\documentclass[aps,prb, twocolumn,showpacs,preprintnumbers,superscriptaddress]{revtex4}

\usepackage[dvipdfm]{graphicx}% Include figure files
\usepackage{dcolumn}% Align table columns on decimal point
\usepackage{bm}% bold math

\begin{document}

%Title of paper

\title{Thermodynamic Properties of Superconducting and Non-Superconducting Pr$_{2}$Ba$_{4}$Cu$_{7}$O$_{15-\delta }$ Compounds with Metallic Double Chains}

\author{Takahisa Konno}
\author{Michiaki Matsukawa} 
\email{matsukawa@iwate-u.ac.jp }
%\homepage[]{Your web page}
%\thanks{}
%\altaffiliation{}
\author{Keisuke Sugawara}
\author{Haruka Taniguchi} 
\author{Junichi Echigoya} 
\affiliation{Department of Materials Science and Engineering, Iwate University, Morioka 020-8551, Japan}
\author{Akiyuki  Matsushita}
\affiliation{National Institute for Materials Science, Ibaraki 305-0047}
\author{Makoto Hagiwara}
%\author{Tsuyoshi Miyazaki}
\affiliation{Kyoto Institute of Technology, Kyoto 606-8585,Japan}
\author{Kazuhiro Sano}
\affiliation{Department of Physics Engineering, Mie University, Tsu 514-8507,Japan}
\author{Yoshiaki Ohno}
\author{Yuh Yamada}
\affiliation{Department of Physics, Niigata University, Niigata 950-2181,Japan}
\author{Takahiko Sasaki}
\author{Yuichiro Hayasaka}
\affiliation{Institute for Materials Research, Tohoku University, Sendai 980-8577,Japan}

%\author{Junichi Echigoya$^{1}$}

\date{\today}

\begin{abstract}
To examine the thermodynamic properties of Pr$_{2}$Ba$_{4}$Cu$_{7}$O$_{15-\delta }$ compounds with metallic CuO double chains, we measured the specific heats of superconducting and non-superconducting  polycrystalline samples  at low temperatures (1.8-40 K) under various magnetic fields (up to 9 T). 
In the as-sintered non-superconducting sample,  a $\lambda-$like enhancement in the specific heat measurement appeared near the antiferromagnetic transition temperature $T_{N}=17$ K. In contrast, the reduced superconducting sample with $T_{c,on}=26.5$ K exhibited no obvious superconducting anomaly in its specific data, but a Schottky-like broad maximum  appeared at low temperatures. 
The Schottky-like anomaly was attributed to low-lying quasi-triplet splitting of  Pr$^{3+}$ ions under the crystal field effect.

\end{abstract}

% insert suggested PACS numbers in braces on next line
\pacs{74.25.Ha,74.25.F-,74.90.+n}
% insert suggested keywords - APS authors don't need to do this
%\keywords{}
\renewcommand{\figurename}{Fig.}
%\maketitle must follow title, authors, abstract, \pacs, and \keywords
\maketitle

\section{INTRODUCTION}

Since the discovery of high-$T_\mathrm{c}$ copper-oxide superconductors, strongly correlated electron systems  have been extensively investigated. Besides the physical properties 
of two-dimensional CuO$_{2}$ planes, researches have focused on the physical role of one-dimensional (1D) CuO chains
in some families of high-$T_\mathrm{c}$ copper oxides.

Structurally, the Pr-based cuprates, PrBa$_{2}$Cu$_{3}$O$_{7-\delta}$ (Pr123) and PrBa$_{2}$Cu$_{4}$O$_{8}$ (Pr124), are identical to their corresponding
Y-based high-$T_\mathrm{c}$  superconductors, YBa$_{2}$Cu$_{3}$O$_{7-\delta}$ (Y123) and  YBa$_{2}$Cu$_{4}$O$_{8}$ (Y124).
Pr123 and Pr124 compounds have insulating CuO$_{2}$ planes and are non-superconductive. \cite{SO87,HO98}  
The suppression of superconductivity in the Pr substitutes has been explained by the hybridization of 
Pr-4$f$ and O-2$p$ orbitals.\cite{FE93}  
The crystal structure of Pr124 with CuO double chains differs from that of Pr123 with CuO single chains. 
It is well known that CuO single chains in Pr123 and CuO double chains in Pr124 show semiconducting and metallic behaviors, respectively.\cite{MI00}
The carrier concentration of doped double chains of Pr124 is difficult to vary,  because it is thermally stable up to high temperatures.
In addition, Pr ions in both Pr123 and Pr124 become antiferromagnetic ordered at the antiferromagnetic transition temperature $T_\mathrm{N}=17$ K.\cite{Li89,YA97} 

\begin{figure}[ht]
\includegraphics[width=8cm]{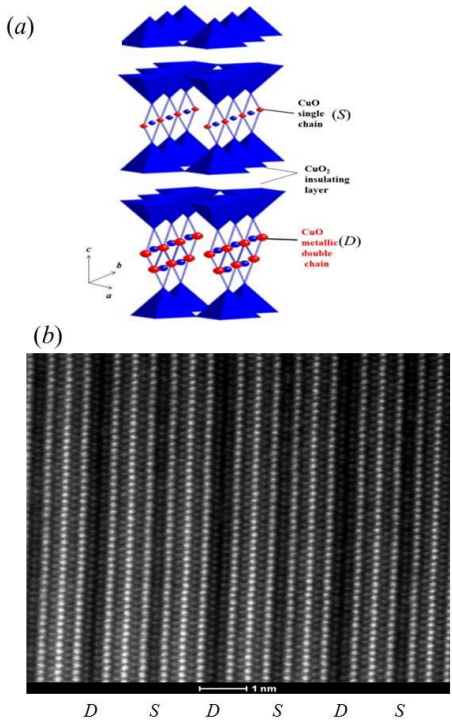}
\caption{(color online) (a)Typical crystal structure of  Pr$_{2}$Ba$_{4}$Cu$_{7}$O$_{15-\delta }$ compound. 
Blocks of CuO single chains (denoted $S$) and CuO double chains ($D$) are alternately stacked along  $c$-axis.  (b) TEM image of superconducting Pr$_{2}$Ba$_{4}$Cu$_{7}$O$_{15-\delta }$ compound with  a superconducting onset temperature $T_\mathrm{c}^\mathrm{on}$ of 26.5 K. 
 }\label{TEM}
\end{figure}

The compound Pr$_{2}$Ba$_{4}$Cu$_{7}$O$_{15-\delta}$ (Pr247) is an  intermediate between  Pr123 and Pr124.   
In this compound,  CuO single-chain and double-chain blocks are alternately stacked along the $c$-axis\cite{BO88,YA94} (see Fig.\ref{TEM}). 
The physical properties of  the metallic CuO double chains can be examined 
by controlling the oxygen content along the semiconducting CuO single chains.
Anisotropic resistivity measurements of single-crystal Pr124 have revealed that metallic transport arises by the conduction along the CuO double chains.\cite{HO00}
In oxygen removed polycrystalline Pr$_{2}$Ba$_{4}$Cu$_{7}$O$_{15-\delta }$, superconductivity appears at an onset temperature $T_\mathrm{c}^\mathrm{on}$ of $ \sim $15 K. \cite{MA04} 
Hall coefficient measurements of  superconducting Pr247 with $T_\mathrm{c}^\mathrm{on}=15$ K have revealed that at intermediate temperatures below 120 K, the main carriers change from holes to electrons, as the temperature decreases. Accordingly, this compound is an electron-doped superconductor.\cite{MA07} 
In our previous study, we examined the effect of magnetic fields on the superconducting phase of Pr247.\cite{CH13}  Despite of the resistive drop associated with the superconducting transition, we found that the diamagnetic signal was strongly suppressed as expected in  the 1D superconductivity of  CuO double chains.  We also reported the temperature dependence of the Hall coefficient in superconducting Pr247 with a higher $T_\mathrm{c}^\mathrm{on}\sim 27$ K.\cite{TA13}
Our findings indicated that the superconducting transition temperature increased because  the density of doped electron carriers became denser under the reduction treatment, consistent with a theoretical prediction.\cite{SA05}

In this paper, we demonstrate the thermodynamic properties of the electron-doped metallic double-chain compound Pr$_{2}$Ba$_{4}$Cu$_{7}$O$_{15-\delta }$, which has a  higher $T_\mathrm{c}^\mathrm{on}$ (26.5 K), under different magnetic fields. 
Figure \ref{TEM} displays the typical crystal structure of Pr$_{2}$Ba$_{4}$Cu$_{7}$O$_{15-\delta }$, in which the CuO metallic double chains and semiconducting single chains are  alternately stacked  along the $c$-axis. 
Section II outlines the experimental methods, and Sec. III presents the magneto-transport and magneto-thermodynamic properties  of polycrystalline Pr247 samples. 
Along with the magneto-transport data, we report the temperature dependences of the specific heats  in the as-sintered and reduced Pr247. These results  are discussed in terms of a three-level system (quasi triplet), in which the ground-state of  the Pr$^{3+}$ ions is split by the crystal field effect.  
The final section is devoted to a summary. 

\section{EXPERIMENT}
Polycrystalline samples of Pr$_{2}$Ba$_{4}$Cu$_{7}$O$_{15-\delta }$(Pr247) were synthesized by the citrate pyrolysis method.\cite{HA06} After several annealing processes,  the resulting precursors were pressed into  a pellet and  calcined at 875-887 $ ^{ \circ }$C for an extended period over 120-180 h under ambient oxygen pressure.
The oxygen in the as-sintered sample was removed by reduction treatment   in a vacuum  at 500 $ ^{ \circ }$C for 48 h,  yielding  a superconducting material. 
Typical dimensions of the pelletized rectangular sample were $9.9\times 2.4\times 1.7$ mm$^{3}$.
The Pr247 sample with $T_\mathrm{c}^\mathrm{on}=26.5$ K was observed by high-resolution transmission electron microscopy  using a JEOL3010 microscope operated at 300 kV at Tohoku University.
The local crystal structure was analyzed from images  obtained by the high-angle annular dark-field scanning transmission electron microscope  method.\cite{PE90} 
The oxygen deficiency in the sample with $T_\mathrm{c}^\mathrm{on}=26.5$ K  prepared by the citrate method was estimated to be  $\delta = 0.56$  from gravimetric analysis. 
As a function of the oxygen deficiency,  the $T_\mathrm{c}^\mathrm{on}$ rises rapidly at $\delta\geq \sim 0.2$, then monotonically increases with increasing $\delta$, and finally saturates  around 26-27 K at  $\delta\geq \sim 0.6$.\cite{HA08}
Accordingly, the carriers in the present sample are concentrated  around the optimally doped region. 

The electric resistivity in zero magnetic field was measured by the $dc$ four-terminal method. The magneto-transport up to 9 T was measured by the $ac$ four-probe method using a physical property measuring system (PPMS, Quantum Design) , increasing the zero-field-cooling (ZFC) temperatures from 2 K  to 40 K. The high field resistivity  (up to 14 T) was measured in a superconducting magnet at the High Field Laboratory for Superconducting Materials, Institute for Materials Research, Tohoku University.  The electric current  $I$ was applied longitudinally to the sample ; consequently, 
the applied magnetic field $H$ was transverse to the sample (because $H\perp I$). 
The specific heats in the ZFC mode were measured to be between 2 K and 40 K by the PPMS. 
For comparison,  the temperature dependence of the specific heat of the superconducting Y$_{2}$Ba$_{4}$Cu$_{7}$O$_{15-\delta }$ (Y247) compound prepared by the high-pressure oxygen method \cite{YA94} was separately measured in zero field at NIMS.  

The $dc$ magnetization was performed under ZFC in a commercial superconducting quantum interference  device magnetometer (Quantum Design, MPMS). Magnetic fields of 0.002 T, 0.01 T, 1 T, 3 T, and 5 T were applied. 

\begin{figure}[ht]
\includegraphics[width=8cm]{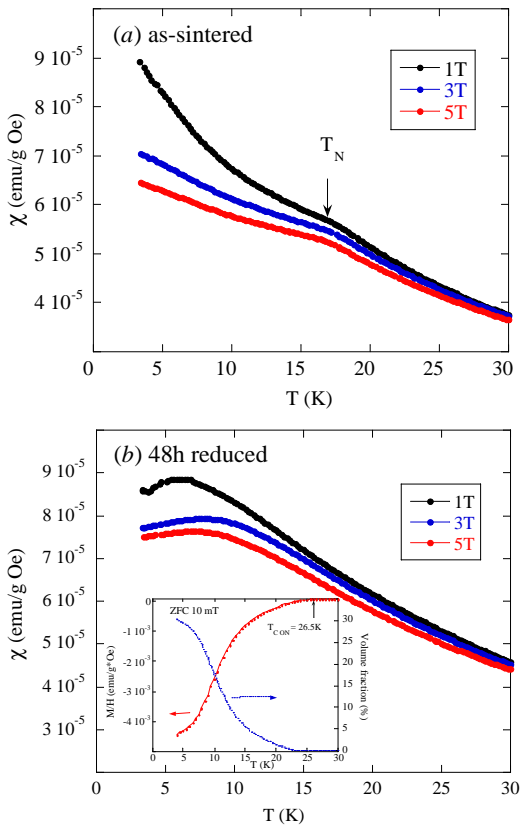}
\caption{(color online) Low-temperature dependences of magnetic susceptibilities  $\chi $ of Pr$_{2}$Ba$_{4}$Cu$_{7}$O$_{15-\delta}$ compounds  under various magnetic fields (1 T, 3 T, and 5 T). (a)Non-superconducting as-sintered sample  and (b) superconducting 48-h-reduced sample  with $T_{c,on}=26.5$ K. 
Inset plots are magnetic data recorded at  10 mT, from which we estimated $T_\mathrm{c}^\mathrm{on}=26.5$ K. Superconducting volume fractions estimated from  present data are also plotted (right vertical axis in inset). 
 }\label{MT}. 
\end{figure}

\begin{figure}[ht]
\includegraphics[width=8cm]{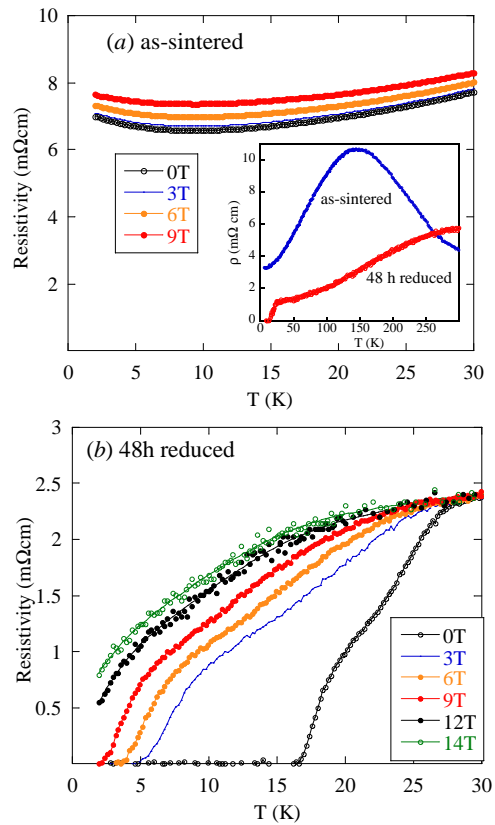}
\caption{(color online)Low-temperature dependences of electric resistivities of Pr$_{2}$Ba$_{4}$Cu$_{7}$O$_{15-\delta}$ compounds measured  under various magnetic fields. (a)  As-sintered non-superconducting sample, measured at  0 T, 3 T, 6 T, and 9 T, and (b)48-h-reduced superconducting sample with $T_{c,on}=26.5$ K, measured at 0 T, 3 T, 6 T, 9 T, 12 T, and 14 T.
For comparison, inset plots resistivity data of both samples versus  $T$ (between 2 K and 300 K). 
 }\label{RT}. 
\end{figure}

\begin{figure}[ht]
\includegraphics[width=8cm]{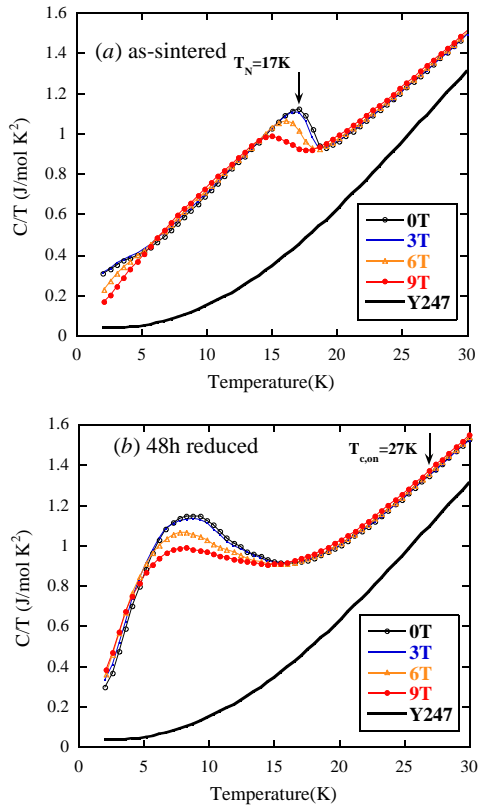}
\caption{(color online) Temperature dependences of specific heats of Pr$_{2}$Ba$_{4}$Cu$_{7}$O$_{15-\delta}$ compounds under various magnetic fields (up to 9 T). (a) The non-superconducting as-sintered sample and (b) the superconducting 48-h-reduced sample  with $T_{c,on}=26.5$ K. For comparison, zero-field specific heat data of superconducting Y$_{2}$Ba$_{4}$Cu$_{7}$O$_{15-\delta }$ (Y247) prepared in high-pressure oxygen are also plotted.  
 }\label{CT}. 
 
\end{figure}

\begin{figure}[ht]
\includegraphics[width=8cm]{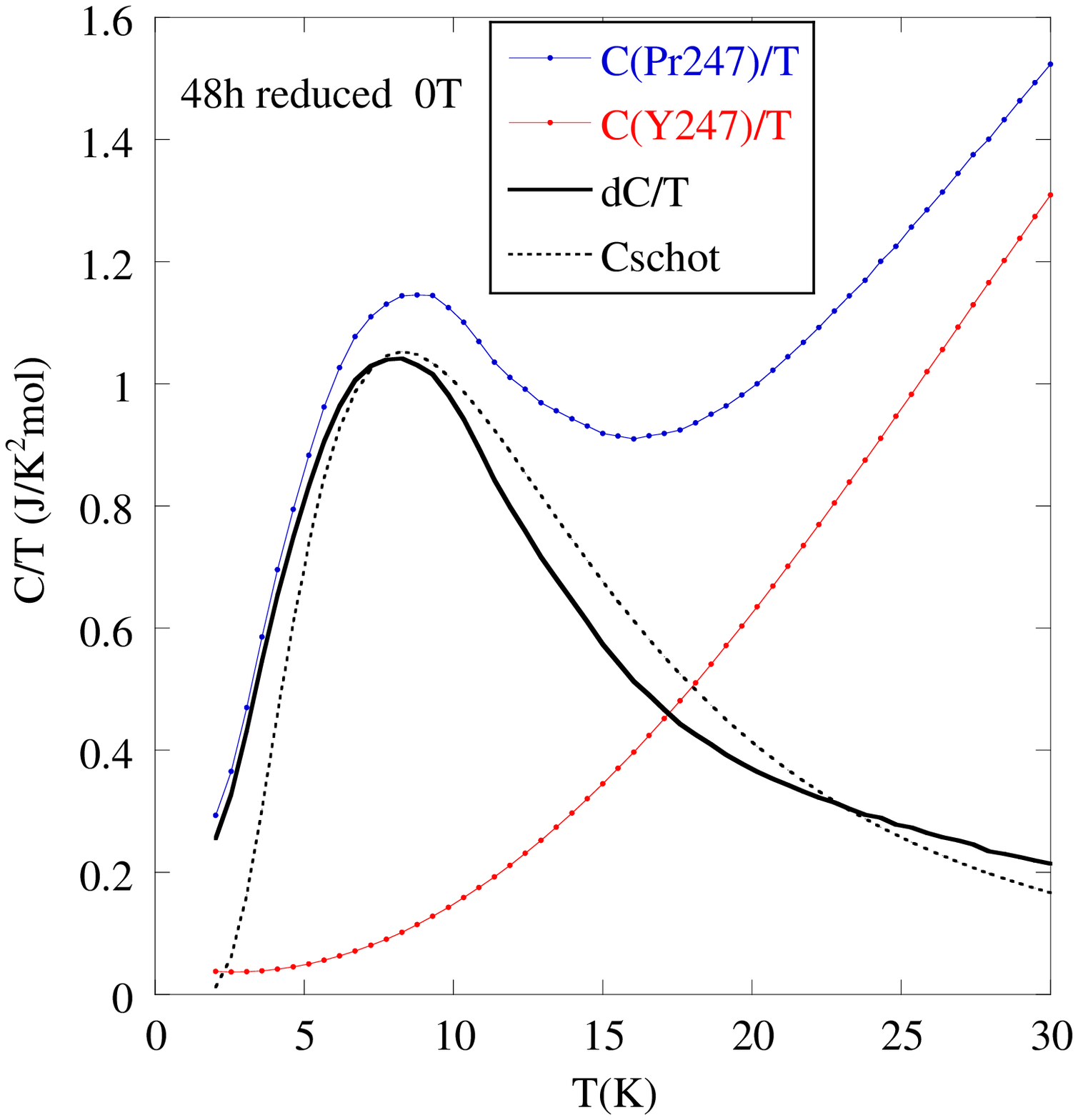}
\caption{(color online) Temperature dependences of specific heat differences $\Delta C/T$ observed in Pr247 and Y247 ($C_{\mathrm{Pr247}}/T$ and  $C_{\mathrm{Y247}}/T$, respectively).
Dashed curves are calculated Schottky  heat anomalies  due to crystal field splitting of Pr$^{3+}$. 
Parameters under zero field were fitted  as $ E_{1}=23.2 K$ and $E_{2}=46.4 K$.
 }\label{CTcal}. 
 
\end{figure}

\section{RESULTS AND DISCUSSION}

Figure \ref{TEM} shows the typical crystal structure and a TEM image of the superconducting Pr$_{2}$Ba$_{4}$Cu$_{7}$O$_{15-\delta }$ compound with  $T_\mathrm{c}^\mathrm{on}=26.5$ K. $S$ and $D$  denote CuO single-chain and double-chain blocks, respectively.
We observe a regular stacking structure of \{-D-S-D-S-D\}.
In the x-ray diffraction pattern of  polycrystalline Pr$_{2}$Ba$_{4}$Cu$_{7}$O$_{15-\delta }$ synthesized by the citrate pyrolysis method, a clear peak corresponding to the Miller index (004) of Pr247 was observed, but peaks of Pr123 or Pr124 phases were absent (data not shown).\cite{CH13}

Figure \ref{MT} plots  the temperature dependences of the magnetic susceptibilities of the non-superconducting and superconducting samples  under various magnetic fields (up to 5 T). In the as-sintered sample, a typical anomaly  in $\chi $ occurs near 17 K.  This anomaly is associated with an antiferromagnetic transition of  the Pr sublattice,  as previously reported in Pr123 and Pr124 systems.\cite{Li89,YA97}  On the contrary, the 48-h-reduced sample becomes superconductive  below  $T_\mathrm{c}^\mathrm{on}=26.5$ K.  The superconducting volume fraction $f$ was estimated  from the magnetic data taken at 10 mT by  $f=\chi\cdot \rho/4\pi \times 100  $, where  $ \chi $ and $ \rho $ denote the magnetic susceptibility per unit weight and the sample density (5.3 g/cm$^3$), respectively. As displayed in the inset of Fig.\ref{MT} (b), the volume fraction reaches $ \sim $30 \% at 4 K, indicating bulk superconductivity. 

However, under high fields, the magnetic data of the superconducting sample monotonically increases as the  temperature decreases to below 17 K and saturate  at low temperatures. This trend reflects the ferromagnetic character of the sample. As highlighted in our previous paper,\cite{CH13}  the diamagnetic signal is suppressed despite the resistive drop associated with
the superconducting transition. This observation is closely related to the 1D superconductivity of  CuO double-chains.  
In contrast to the positive magnetization data at higher field,  
 the resistivity data of the 48-h-reduced sample substantially drop as the transport currents becomes superconducting  
(see Fig.\ref{RT}(b)). 
The clear difference in the magnetic behavior between the as-sintered and reduced samples reflects the different magnetic interactions among the Pr magnetic ions. 
The Pr atoms in Pr124,  are well separated, so their  antiferromagnetic ordering is not dominated by the superexchange interaction.  Instead, we consider that the antiferromagnetic coupling of the Pr sublattice occurs due to the Rudermann-Kittel-Kasuya-Yoshida (RKKY) interaction and is mediated by itinerant carriers as noted by Xu et al.\cite{XU10}  In Pr247, the vacuum reduction treatment varies the distance between the Pr ions, and the ferromagnetic property of the reduced sample occurs through the RKKY interaction. 

From the magnetic susceptibility measurements over a wide range of temperatures range (20-200 K), we estimate the effective magnetic moment  $\mu _\mathrm{eff}$ of the Pr ions   by the Curie-Weiss law. Performing the calculation, we obtained that  $\mu _\mathrm{eff}$=3.02 and 3.26 $\mu _\mathrm{B}$ for the non-superconducting and superconducting samples, respectively. 
These values are almost consistent with the effective moment of Pr124 ($\mu _\mathrm{eff}=3.11 \mu _\mathrm{B}$).\cite{YA97} In particular, the $\mu _\mathrm{eff}$ of the  superconducting sample is close to that of Pr$^{3+}$(3.54 $\mu _\mathrm{B}$). 

The magnetoresistance of the as-sintered and 48-h-reduced samples also differ at high temperatures (above $T_\mathrm{c}^\mathrm{on}=26.5$ K).
In our next paper, concerning the effects of external pressure  on the magneto-transport of superconducting and non-superconducting Pr247, we will link 
this difference to the  electronic phase diagram  of the 1D zigzag CuO chain model.

Figure \ref{CT} plots the temperature dependences of the specific heats $C/T$ of the  non-superconducting and superconducting Pr247 compounds under several magnetic fields (up to 9 T). The $C/T$  of the non-superconducting as-sintered sample exhibits a $\lambda-$ like enhancement associated with the antiferromagnetic transition of Pr ions, as previously reported in  non-superconducting Pr123 and Pr124 systems.\cite{KE89, YA97} 
As the external magnetic field increases, this sharp increase is suppressed and peaks at lower temperatures. 
At 9 T, the thermodynamic peak is considerably depressed  and is located around 15 K.  At low temperatures (below 5 K), the specific heat also exhibits a strong field dependence, for reasons which have not yet been clarified. 
On the other hand, the   $C/T$  of the superconducting sample is not enhanced at temperatures above 15 K, regardless of the applied magnetic field.  
The absence of $\lambda-$ like enhancement in reduced Pr247 appears to be  consistent with the corresponding magnetic behavior, which shows no  clear AFM transition.  
At temperatures below 15 K,  the $C/T$ of the superconducting sample shows a broad peak around 8.5 K.  
 This broad anomaly is strongly suppressed under a  magnetic field.  
For comparison,  Fig.\ref{CT} also plots  the temperature dependence of the specific heat of superconducting Y247 in the absence of magnetic field, which was prepared by the high pressure oxygen method.   
Oxygen removed Y$_{2}$Ba$_{4}$Cu$_{7}$O$_{14}$, with  no superconductivity down to 2 K, exhibits a   $C/T$  magnitude and temperature dependence similar to those of  of the superconductor Y247. \cite{IR02}
Accordingly, because Y247 and Pr247 share the same crystal structure, we roughly assumed the low-temperature specific heat of superconducting Y247 to be  the lattice component of the Pr247 sample.  We further assumed that the low-temperature dependences of the electronic components of both Y247 and Pr247  were significantly smaller than those of  the lattice parts.

Figure \ref{CTcal} plots  the specific heat differences $dC/T$ between the superconducting Pr247 and Y247 data,  ($C_{\mathrm{Pr247}}/T$ and  $C_{\mathrm{Y247}}/T$, respectively) versus temperature. 
%The dashed curve indicates the Schottky  anomaly arising from a thermal population of electrons  on the basis of a low-lying quasi-triplet splitting of  Pr$^{3+}$ ions due to the crystal field effect.
Comparing our results with the previous data obtained using Pr-based 123,\cite{HI94}  it is evident that the low-temperature specific heat data of  PrBa$_{2}$Cu$_{3}$O$_{7}$  and  PrBa$_{2}$Cu$_{3}$O$_{6}$  follow temperature trends, similar to  those of as-sintered and vacuum-reduced  Pr247, respectively.  
The common features include the $\lambda $-like enhancement in the former compounds and the broad  $C/T$ maximum in the latter. 
To analyze the low-temperature peak  in the specific heat trends of PrBa$_{2}$Cu$_{3}$O$_{6}$, Hilscher et al\cite{HI94} introduced a Schottky type specific heat contributed by  a thermal population of electrons. In this study, we followed Hischer's approach to understand  the broad maximum  in the  $C/T$  of  the corresponding superconducting Pr247.\cite{KI76}   
 In the Pr123 system, the crystal field ground state of the Pr ion is  a quasitriplet and is separated by several tens of meV from the group of remaining crystal field levels.\cite{SO91,HI94}  Accordingly, we assumed that at temperatures below 30 K,  the energy levels of the Pr ion  are split into three by the crystal field effect. 
 
Assuming that for three energy levels $E_{i} (i=0,1, \mathrm{and}\ 2)$  $E_{0}$=0 K and $E_{1}<E_{2}$,  we then get a partition function for the three energy level system, $Z=1+e^{-\beta E_{1}}+ e^{-\beta E_{2}} $  ($\beta =1/kT$). 
Substituting the above formula into a general expression for a specific heat, $C=\frac{1}{kT^2}\frac{d^2\log Z}{d\beta ^2}$, 
 a Schottky-type  specific heat  with three energy levels is given by

 \[C_{\mathrm{Sch}}= k_\mathrm{B}\frac{ x^2e^{-x}+y^2e^{-y}+(x-y)^2e^{-x-y}}{(e^{-x}+e^{-y}+1)^2}  \]
where $x=E_{1}/T$ and $y=E_{2}/T$  denote the temperature reduced energy levels.
% If it is assumed that $E_{0}$=0 and $E_{1}<E_{2}$,  $E_{1}$ and $E_{2}$ then represent the energy splitting due to the crystal field effect. 

As shown in Fig.\ref{CTcal}, the zero field data  below 30 K are roughly fitted by  the  Schottky  expression with  $E_{1}=23.2 K$ and $E_{2}=46.4 K$. 
The energy levels fitted to the curves of the superconducting Pr247 compound are similar to the experimental and calculated crystal field levels of the $^{3}H_{4}$ multiplet in PrBa$_{2}$Cu$_{3}$O$_{6}$, which has orthorhombic crystal field parameters.
For example,  as predicted from inelastic neutron scattering measurements of PrBa$_{2}$Cu$_{3}$O$_{6}$, the crystal field splitting  of this compound is  $E_{1}=19.7 K$ and $E_{2}=39.4 K$. \cite{HI94} 
The low-temperature peak in the specific heat of the oxygen-removed superconducting sample is reasonably described by the Schottky  anomaly
(see Fig.\ref{CTcal}).
The small discrepancy between the  calculated curve and the experiment data is probably attributed to magnetic interactions among the Pr ions. 
The reduced superconducting sample shows no obvious anomaly associated with the superconducting transition, inconsistent with  the experimentally observed resistive drop and diamagnetic signal as the temperature decreases.  
As shown in the inset of Fig.\ref{MT}, the superconducting volume fraction is  several percent around 15 K, and rapidly increases as the temperature falls below 10 K.  
Thus, we infer that  the superconducting anomaly of metallic CuO double chains is masked by  the Schottky like broad enhancement in  $C/T$ caused by  the crystal field effect of the Pr ions.

\section{SUMMARY}
We demonstrated the thermodynamic properties of superconducting and non-superconducting Pr$_{2}$Ba$_{4}$Cu$_{7}$O$_{15-\delta }$ compounds with metallic CuO double chains.
For this purpose, we measured the specific heats  of the  polycrystalline samples  at low temperatures under varying  magnetic fields (up to 9 T).  
The specific heat of the as-sintered non-superconducting sample displayed  the $\lambda-$like enhancement near the antiferromagnetic transition temperature $T_{N}=17$ K. 
This anomaly around the superconducting transition was absent in the reduced superconducting sample  
for reasons related to the small superconducting volume fraction at temperatures above 15 K. As  the temperature decreased below 15 K, the  $C/T$ of the superconducting sample broadly peaked around 8.5 K.  
This Schottky-like broad maximum  was attributed to  low-lying quasi-triplet splitting of  Pr$^{3+}$ ions under  the crystal field effect.

%X-ray powder diffraction pattern of nominal Pr$_{2}$Ba$_{4}$Cu$_{7}$O$_{15-\delta }$ reveals the coexistence of Pr123 and Pr124 phases.  
\begin{acknowledgments}
The authors are grateful for  M. Nakamura for his assistance in PPMS experiments at Center for Regional Collaboration in Research and Education, Iwate University. 
%They thank Dr. A. Matsushita for his collaboration in Hall coefficient measurement. 
%This work was partially supported by a Grant-in-Aid for Scientific Research from Japan Society of the Promotion of Science. 
\end{acknowledgments}

\end{document}